 \documentclass[pre,floatfix,amsmath,amssymb,
    preprintnumbers,
    a4paper,
    showpacs,
    twocolumn,
    ]{revtex4}

 \usepackage{graphicx}

\def\[{\begin{eqnarray}}
\def\]{\end{eqnarray}}
\def\beg{\begin{eqnarray}}
\def\ende{\end{eqnarray}}
\renewcommand{\vec}[1]{\textbf{#1}}

\def\qqq{\qquad\qquad\qquad}

\newcommand{\rot}[1]{\mathop{\rm rot}\nolimits #1}
\renewcommand{\div}[1]{\mathop{\rm div}\nolimits #1}
\def\unitr{\hat{\mathbf{e}}_R}
\def\unitp{\hat{\mathbf{e}}_\phi}
\def\unitz{\hat{\mathbf{e}}_z}
\def\Rin{R_\mathrm{in}}
\def\Rout{R_\mathrm{out}}
\def\Omin{\Omega_{\mathrm{in}}}
\def\Omout{\Omega_{\mathrm{out}}}
\def\muh{\hat{\mu}}
\def\etah{\hat{\eta}}
\def\Omend{\Omega_{\mathrm{end}}}

\def\Omegab{\Omega_\mathrm{0}}
\def\Omegad{\Omega_\mathrm{D}}
\def\Rey{\mathrm{Re}}

\def\Bext{\vec{B}_0}

\def\Pm{\mathrm{Pm}}
\def\Ha{\mathrm{Ha}}
\def\mperm{\mu_0}
\def\mdiff{\eta}

\def\epslayerk{\epsilon} 
\def\omepsilon{\varepsilon} 

\def\ekdep{d_\mathrm{E}}
\def\hadep{d_\mathrm{H}}

\def\dpump{\beta}
\def\dpumpc{\beta_c}
\def\raydisc{\zeta}
\def\stdwid{9cm}
\def\Omplain{\Omega_{\mathrm{plate}}}
\def\Omfluid{\Omega_{\mathrm{fluid}}}

\newcommand{\ord}[1]{$O(#1)$}
\newcommand{\nablasq}{\nabla^2}
\def\pre{Phys.Rev.E}

\begin{document}

\title{The Ekman-Hartmann layer in MHD Taylor-Couette flow}

\author{Jacek Szklarski, G\"unther R\"udiger}
\affiliation{Astrophysikalisches Institut Potsdam,
             An der Sternwarte 16, D-14482 Potsdam, Germany}
\email{jszklarski@aip.de} 
\date{\today}

\begin{abstract}
We study magnetic effects induced by rigidly rotating plates enclosing a
cylindrical MHD Taylor-Couette flow at the finite aspect ratio $H/D=10$.
The fluid confined between the cylinders is assumed to be liquid metal
characterized by small magnetic Prandtl number, the cylinders are
perfectly conducting, an axial magnetic field is imposed $\Ha \approx 10$, 
the rotation rates correspond to $\Rey$ of order $10^2-10^3$.
We show that the end-plates introduce, besides
the well known Ekman circulation, similar magnetic effects which arise
for infinite, rotating plates, horizontally unbounded by any walls.
In particular there exists the Hartmann current which penetrates the
fluid, turns into the radial direction and together with the applied
magnetic field gives rise to a force.  Consequently the flow can be compared 
with a Taylor-Dean flow driven by an azimuthal pressure gradient.
We analyze stability of such flows and show that the currents
induced by the plates can give rise to instability for the considered
parameters.  When designing an MHD Taylor-Couette experiment, a special
care must be taken concerning the vertical magnetic boundaries so they do not
significantly alter the rotational profile. 
\end{abstract}


\maketitle

\section{Introduction}

Motion of a fluid confined between two concentric, rotating cylinders
is a classical problem in hydrodynamics and, if the fluid is conducting
and an external magnetic field is applied, magnetohydrodynamics (MHD).
The flow of this type, usually referred to as the Taylor-Couette flow,
has been first studied by Couette~\citep{Couette1888} and later was
subject of a seminal work by Taylor~\citep{Taylor1923}, who experimentally
confirmed theoretical results of a linear stability analysis.  In the
field of MHD, an important work was done by Velikhov~\citep{Velikhov59}
who has shown that for the conducting fluid a weak magnetic field can
play a destabilizing role and can lead to an instability which today is
called magnetorotational instability (MRI~\citep{Jietal01}).

When studying the Taylor-Couette system it is common to assume some
simplifications, the small gap approximation or large aspect ratio.
In the former it is assumed that the gap between the cylinders $D = \Rout-\Rin$ 
is small compared to the radii, i.e., $D/\Rout \ll 1$,
this allows the neglect of terms of order $1/R$, $R$ being distance
from the center of rotation.  When considering the large aspect ratio,
one assumes that the height of the cylinders $H$ is much larger than
the gap width $\Gamma=H/D \gg 1$, which guarantees that a secondary flow
due to the plates bounding the cylinders is insignificant and does not
disturb the rotational profile of the fluid.

On the other hand, there is also plenty of work done for small aspect
ratio $\Gamma \approx 1$, where the rigidly rotating end-plates play
crucial role and simply introduce a new class of problems.  When $\Gamma$
becomes an important parameter it is possible to observe a wide family of
different states (including non-axisymmetric ones or peculiar asymmetric
patterns -- anomalous modes) for the same parameters, so that the
observed results depend on their path through the parameters space
from an initial state.  Therefore this system is an excellent subject
to the bifurcation theory~\citep{1988JFM...191....1P, LopezMarques03,
Mullinetal02, Furukawaetal02, Kageyamaetal04, Youd06}.

In the present work we focus on the case of wide gap $\Rin/\Rout=1/2$
and $\Gamma=10$ which is an intermediate aspect ratio, between very
short and long containers, yet in purely hydrodynamical contest the
influence of the vertical boundaries is small, at least for Reynolds
numbers of order \ord{10^2-10^3}.  However, if the rotation rates are
large enough, so that the corresponding Reynolds number is \ord{10^5} and
larger, the plates can easily dominate the flow in the entire container.
This is due to the Taylor-Proudman theorem, from which follows that in
rapidly rotating systems the flow tends to align itself along the axis
of rotation.  For such rotations it is necessary that $\Gamma$ would have
to be several thousand in order to obtain the rotational profile which
is not profoundly altered by the end-plates~\citep{HollerbachFournier04}.

Results of a recent MRI experiment PROMISE~\citep{MRIexpA, MRIexpB, Stefani07}, 
as well as nonlinear simulations~\citep{SzklRud06,Szklarski07} 
indicate that for a flow with relatively small Reynolds
number $ \approx 10^3$, and parameters resembling essentially MHD stable
flow in the limit of infinitely long cylinders, there exist unexpected
time-dependent fluctuation of the velocity field.  These disturbances
arise as an effect of the vertical boundary conditions, moreover the
simulations show that they are much stronger if the end-plates bounding
the cylinders are assumed to be perfectly conducting.

The plates induce a well known hydrodynamical effect -- the Ekman
circulation, which is a result of unbalanced pressure gradients in
vicinity of the vertical no-slip boundary conditions.  There  the Ekman
layer develops  in which the fluid velocity from the bulk of the container
must match the velocity imposed by the end-plates.

It seems that for MHD Taylor-Couette flow, magnetic effects, unlike
the classical hydrodynamical Ekman layer, induced by the plates have
been overlooked. In this paper we argue that the rigidly rotating plates
together with an imposed axial magnetic field give rise to a similar layer
which develops for an infinite, rotating plate serving as a boundary for
the conducting fluid.  One of the most important features of such flow
is the existence of the Hartmann current (absent in the conventional
Hartmann problem,~\citep{2004JFM...504..183K}) which leaves the boundary
layer and then interacts with the magnetic field.
 In particular, this becomes important for conducting plates which was
the case for the PROMISE experiment, since one of the end-plates was
made from copper.

We discuss properties of Ekman-Hartmann layers for infinite, rotating
plates and relate it to the end-plates enclosing the cylinders in a
Taylor-Couette setup.  It is shown that for considered radial boundary
conditions, the induced current turn eventually in the radial direction
and acting in concert with the imposed axial magnetic field gives rise
to a body force.

We demonstrate that magnetic effects induced by the end-plates enclosing
the cylinders can profoundly alter flow properties. In particular the
rotational profile can become significantly different from the expected
parabolic Couette solution.  Moreover, if the Hartmann current is strong
enough, it is likely that the local Rayleigh criterion for stability
will be violated and the flow becomes centrifugally unstable.  In an MRI
experiment it is crucial to rule out such instabilities and a special
care concerning the vertical boundary conditions is needed in order to
obtain the desired rotational profile.

\section{Problem formulation}
\def\pregrad{\partial_\phi p}

We consider two concentric cylinders with radii $\Rin, \Rout$
embedded in an external axial magnetic field.  They rotate with angular
velocities $\Omin, \Omout$, the radius ratio is $\etah=\Rin/\Rout$, the
rotation ratio $\muh=\Omout/\Omin$. Cylindrical coordinates
$(R,\phi,z)$ with unit vectors $\unitr, \unitp, \unitz$ are used.  
If the cylinders are unbounded, i.e., infinitely
long or periodic, the rotational profile is
 \[ 
   \label{eq:couprofile}
   \Omegab(R)=a+\frac{b}{R^2},
 \]
with
 \[
   a=\Omin \frac{\muh-\etah^2}{1-\etah^2} \quad b=\frac{1-\muh}{1-\etah^2}\Rin^2 \Omin,
 \]
and $u_R=u_z=0$ everywhere.  The flow is hydrodynamically stable if the
Rayleigh criterion $d(R^2 \Omega)^2/dR > 0$ is fulfilled, i.e., for $\muh > \etah^2$.  
Consequently, for the considered radius ratio $\etah=1/2$,
the flow is always stable if $\muh > 0.25$.  Here we consider
only cases when $\muh > 0.25$ so that hydrodynamical instabilities are
ruled out.

Let us introduce the Reynolds number $\Rey$, which measures the rotation rates, and
the Hartmann number $\Ha$, which measures the strength of the externally applied 
magnetic field $\Bext=B_0 \unitz$,
 \[
 \Ha = B_0 \sqrt{\frac{D^2}{\mperm \rho \nu \mdiff}}, 
       \quad \Rey = \frac{\Omin \Rin D}{\nu},
 \]
where $\rho$ is the density, $\nu$ the kinematic viscosity, $\eta$
is the magnetic diffusivity, $\mperm$ is the magnetic
permeability. The fluid confined between the cylinders is assumed to
be incompressible and it can be characterized by the magnetic Prandtl
number $\Pm = \nu/\mdiff$.  For laboratory liquid metals, like gallium,
$\Pm$ is very small -- of order $10^{-(5 \dots 6)}$, therefore 
we concentrate on effects arising only when $\Pm$ is small.

\subsection{The Equations}

Using $D$ as the unit of length, $\nu/D$ as the unit of velocity, $D^2/\nu$ as
the unit of time, $B_0$ as the unit of the axial magnetic field and assuming $\vec{B} = \Bext + \vec{b}$
we can write non-dimensional MHD equations for the problem of our interest, i.e.
 \begin{subequations}
 \label{eq:mhd}
 \[
  \lefteqn{ \partial_t \vec{u} + (\vec{u} \cdot \nabla)\vec{u} = -\nabla p + \nablasq \vec{u} }
	    \nonumber \\
	    && \quad \quad \quad   + 
	       \frac{\Ha^2}{\Pm} 
              \left[ (\rot{\vec{b}}) \times \vec{b} 
                +(\rot{\vec{b}}) \times \Bext/B_0 \right], \ \ \ \ \ \ \  \label{eqn-momentum}\\
  \lefteqn{  \partial_t \vec{b} =
                  \frac{1}{\Pm} \nablasq \vec{b} 
		 + \rot{(\vec{u} \times \vec{b})}
		 + \rot{(\vec{u} \times \Bext/B_0)}, } \label{eqn-induc} 
 \]
 \end{subequations}
with $\div{\vec{u}} =  \div{\vec{b}} = 0$, where $\vec{u}$ and $\vec{b}$ are the
velocity and the perturbed magnetic field, $p$ is the pressure. 

For the velocity we apply no-slip boundary conditions at the cylinders 
and at the end-plates as well.  We assume that both the plates rotate rigidly with 
angular velocity $\Omend$, which can be set to any value so that the plates
can rotate independently of the cylinders. 

Boundary conditions for the magnetic field are determined by magnetic properties of
the cylinders and the plates.  Here we consider only perfectly conducting radial boundaries,
so that the transverse currents and perpendicular component of the magnetic field vanish, 
hence $R^{-1} b_\phi + \partial_R b_\phi = 0$ at $R=\Rin/D, R=\Rout/D$.  
We chose such boundaries since in the PROMISE experiment the cylinders were made
of copper.  The reason for choosing copper is that 
the critical $\Rey$ and $\Ha$ numbers for the onset of the 
MRI are smaller by almost a factor of 2 with perfectly conducting boundaries than with 
insulating boundaries \citep{RudAN05}.

For the end-plates, similarly like for the 
walls, the electric field must be continuous and $b_z=0$,  
then $b_\phi = \epslayerk \partial_z b_\phi$ at $z=0$ and 
$b_\phi = -\epslayerk \partial_z b_\phi$ at $z=\Gamma$ where $\epslayerk$ characterizes 
a thin layer of relative conductance of the fluid and the plates~\citep{Loper70, HollerbachEps}. 
When $\epslayerk \rightarrow 0$ we obtain conditions corresponding to insulating 
end-plates, i.e., $b_\phi = \partial_z j_\phi = 0$ at $z=0, z=\Gamma$, $j_\phi$ being the azimuthal 
current. For $\epslayerk \rightarrow \infty$ we have the case 
describing the perfectly conducting plates, $\partial_z b_\phi = j_\phi = 0$ at $z=0, z=\Gamma$.
We note that this thin-wall approximation is valid only when the magnetic field varies
linearly within the plates and it does not necessarily resembles situation in a real experiment.

\subsection{The small Pm limit}

 For laboratory liquids the conductivity $\sigma$ is small, so that the
 magnetic diffusivity $\mdiff = 1/\mperm \sigma$ is very large (compared
 to the viscosity) and the corresponding magnetic Prandtl number $\Pm$
 is small.  Consequently the time scale for magnetic diffusion is
 much shorter than other time scales.  Therefore we consider the limit
 $\mdiff \rightarrow \infty$, however it must be supposed that $\Ha$
 tends to a finite value.  The perturbations $\vec{b}$ of the externally
 applied field induced by the motion of the fluid are $\Pm$ times smaller
 than $\Bext$, although theirs effect on the Lorentz force can not be
 neglected since $\Ha^2/\Pm [(\rot{\vec{b}}) \times$$\Bext/B_0]$ is
 already of order $\Ha$. Nevertheless the interactions 
 $(\rot{\vec{b}}) \times \vec{b}$ are vanishingly small.

 Similarly in the induction equation we may apply a quasi-static
 approximation, so that the electromagnetic field proceeds along a
 sequence of steady-state solutions of the Maxwell equations to conditions
 described by $\vec{u}$, and  therefore $\vec{b}$ in each moment adjusts
 instantaneously to the velocity $\vec{u}$.  Hence, in the small Prandtl
 limit $\Pm \rightarrow 0$, the system \eqref{eq:mhd} can be written as
 \begin{subequations}
   \label{eq:quasimhd}
   \[
     \lefteqn{ \partial_t \vec{u} + (\vec{u} \cdot \nabla)\vec{u} = 
       -\nabla p + \nablasq \vec{u} + }
       \nonumber \\
       && \qquad \qqq \Ha^2 (\nabla \times \vec{b}) \times \Bext/B_0, \ \ \ \ \ \  \label{eq:momentumB}\\
     \lefteqn{ \nablasq \vec{b} = - \nabla \times (\vec{u} \times \Bext/B_0) } \label{eq:inducB} 
   \]
 \end{subequations}
 with $\div{\vec{u}} = \div{\vec{b}} = 0$,~\citep{Roberts67, ZikanovThess98}.  
 The equations~\eqref{eq:quasimhd} together with
 the discussed boundary conditions are solved with finite difference
 method using stream function-vorticity formulation in the $(R,z)$ plane.
 In this work we assume that the flow is axisymmetric.  For more details
 on the numerical procedure see~\citep{SzklRud06, Youd06}.

\section{\label{sec:ehlayer}The Ekman-Hartmann layer in the MHD Taylor-Couette flow}

At an interface between an incompressible fluid with low viscosity and
a rapidly rotating rigid surface develops an Ekman layer with thickness
$\ekdep \propto \sqrt{\nu/\Omega}$, where $\Omega$ is rate of uniform
rotation.  Similarly for a flow of conducting, incompressible fluid
in vicinity of a rigid non-rotating boundary, and under the influence
of an external magnetic field perpendicular to the surface there
exists a Hartman layer with thickness $\hadep \propto \Ha^{-1}$. When
these two effects are combined, the Ekman-Hartmann layer develops~\citep{AchesonHide73}.  
It can be viewed either as a modification of
the Ekman layer by introducing the conducting fluid and imposing the
external magnetic field or as a modification of the Hartmann layer by
adding the uniform rotation of the bounding surface.  The resulting layer
(in its steady form) assures a proper transition for the velocity and the
magnetic field from values inside the bulk of the fluid to the applied
boundary conditions.

The linear analysis of the Ekman-Hartmann layer in its idealized case was
presented by~\citet{GilmanBenton68}. They have considered an infinite,
insulating plate  rotating with $\Omplain$ at $z=0$, and a conducting
fluid filling the space $z>0$, the fluid far from the plate rotates with
$\Omfluid=\Omplain(1+\omepsilon)$, $\omepsilon \ll 1$. The most important
conclusion of this work was that in addition to the well known Ekman
suction/blowing of mass flux there also exists an electric Hartmann
current which has the same direction (or opposite when the external
$B_z$ is negative) as the velocity of the Ekman blowing (that if fluid
is blown away or sucked towards the boundaries depends only on sign
of $\omepsilon$).  This current, which is arising due to the vertical
shears, leaves the Ekman-Hartmann layer and potentially influences the
flow far away from the boundary.

\begin{figure*}[!hbt]    
 \mbox{ \includegraphics[width=\stdwid]{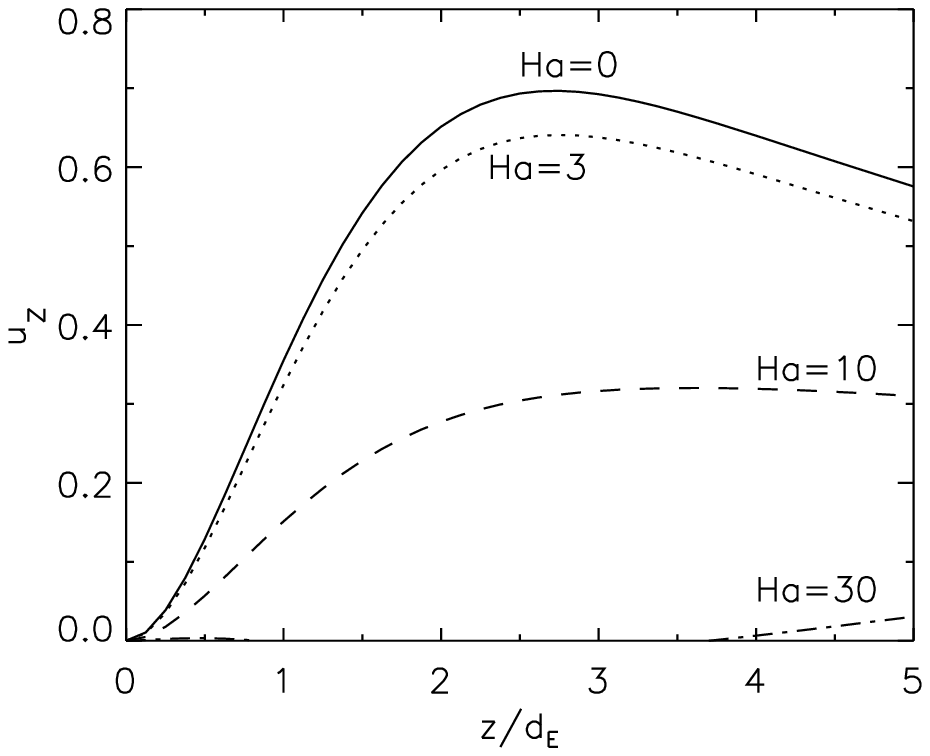}
 \includegraphics[width=\stdwid]{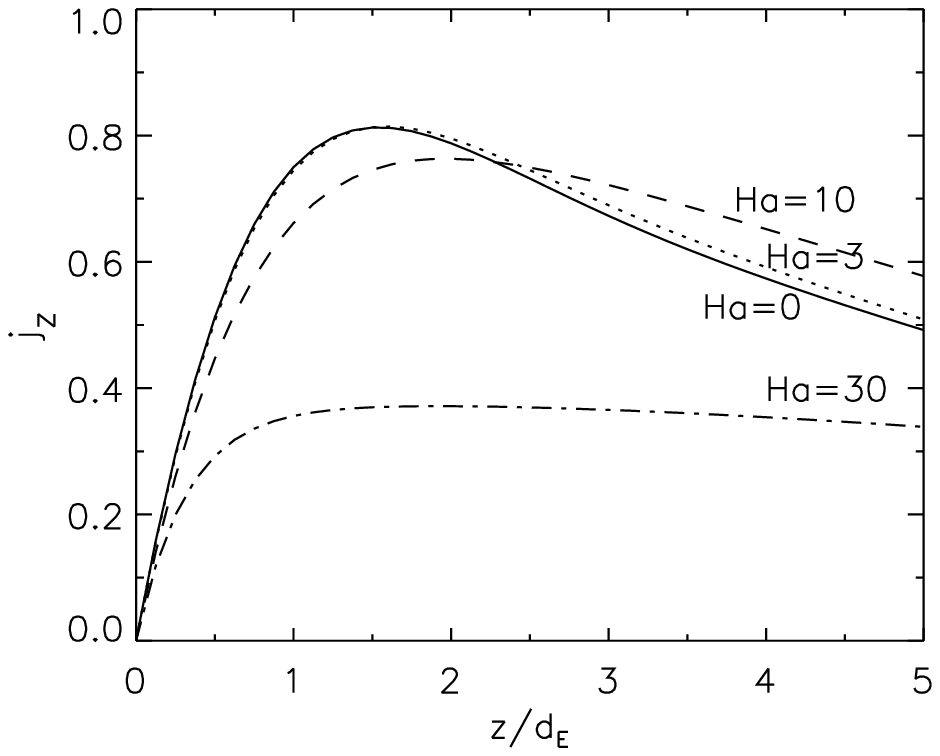} }
 \caption{ \label{fig:ehstruct}
         Structure of the Ekman-Hartmann layer for the MHD Taylor-Couette flow with 
         $\Omin=\Omout=100$, $\Gamma=10$, insulting end-plates rotate with $\Omend=90$,
	 $z$ is the distance from one of the plates. 
	 Different values of the magnetic interaction parameters correspond to: 
         $\Ha \rightarrow 0.0\ (\alpha \rightarrow 0.0)$,
         $\Ha=3.0\ (\alpha=0.2)$,  $\Ha=10.0\ (\alpha=0.7)$,  $\Ha=30.0\ (\alpha=2.1)$, 
         $\ekdep=0.01 D$ is the Ekman layer thickness. 
	 }
\end{figure*}

For the magnetized Taylor-Couette with finite aspect ratio, i.e., if the
cylinders are covered with rigidly rotating end-plates (insulating or
conducting), the Ekman-Hartmann layer also develops.  Naturally influence
of the vertical walls introduces additional important effects and direct
quantitative comparison with the previous work is not possible.  We must
take into account that the fluid which was ejected due to the Ekman
blowing mechanism must eventually get back due to the conservation of
mass and finiteness of the container.  Nevertheless we will show that
the rotating end-plates induce the Hartmann current which can change
the global properties of the flow.

Let us introduce the parameter $\alpha$
which measures the overall importance of the magnetic field,
 \[ \label{eqn-alpha} 
    \alpha = \frac{\ekdep}{\sqrt{2}\hadep} = \frac{\Ha}{\sqrt{2 \Rey \muh}}, 
 \]
where $\hadep = D \Ha^{-1}$ is the Hartmann depth.  For the Ekman depth,
as a measure of the uniform rotation we use $\Omout$.  The magnetic
effects start to be significant when $\alpha \gtrsim 1$, in the limit
$\alpha \rightarrow 0$ we have the classical Ekman layer and for 
$\alpha \rightarrow \infty$ the classical Hartmann layer.  We notice that for
slow rotation corresponding to $\Rey$ of order \ord{10^2-10^3} and
$\Ha$ of order \ord{10}, $\alpha \approx 1$, and therefore we expect
the magnetic fields to be important for many laboratory experiments.

\subsection{Insulating end-plates}

First we consider a case when both the cylinders rotate with the same
angular velocity $\Omin = \Omout = 100$, i.e., $\muh = 1.0$, and the
rotational profile \eqref{eq:couprofile} is flat.  The aspect ratio
is $\Gamma=10$ and the insulating plates rotate with angular velocity
slightly different than the cylinders, $\Omend = 90$.

Figure~\ref{fig:ehstruct} shows how the axial velocity $u_z$ and the
axial current $j_z$ change with distance $z$ from the plates, for
different strength of the applied magnetic field.  It can be seen that
the axial velocity and the axial current decrease for stronger magnetic
field. The explanation is as follows.  The vertical shears in $u_R$
and $u_z$ produce currents which together with axial field generate
body forces acting against the shears.  Since the radial flow must
vanish at the boundaries as well as it vanishes far away from them,
the effect is to reduce the $u_R$ and, due to mass conservation, $u_z$.
Therefore the external axial magnetic field inhibits the Ekman blowing
(which is completely suppressed when $\alpha \rightarrow \infty$) and
makes the boundary layer thinner.  The azimuthal flow $u_\phi$, on the
other hand, is forced to have different values at the boundaries and
far away from them, thus the shear can be decreased only in the region
close to the boundary.

These results are in a good agreement with the linear
solution~\citep{GilmanBenton68} for the case of the infinite, rotating
plate and $\Pm \rightarrow 0$ (the agreement for other quantities like
$u_R$, $u_\phi$ is pleasing as well).  We notice that in the radially
unbounded case. $u_z$ and $j_z$ are independent of $R$ which can not be
true for the enclosed Taylor-Couette system.  The values presented in
Fig.~\ref{fig:ehstruct} are computed for $R=\Rin+D/2$, in the middle of
the gap, so that the influence of the rotating cylinders is smallest.

We point out that the induced axial current $j_z$ (the Hartmann current)
exists outside the boundary layer.  This is not the case for non-rotating
Hartmann boundaries.  For unbounded flow this current quickly converges
to an asymptotic constant value, but for the case of flow between two
plates, or for the enclosed cylinders, it can not be true and currents
induced by both end-plates must eventually interact.  When we consider a
system symmetric in the $z$ direction, i.e., when the two plates rotate
in the same manner, the induced $j_z$ have the same strength but opposite
signs and they eventually meet turning into the radial direction  (and
consequently $j_z=0$ in the middle of the container for the symmetric
boundary conditions).

We have varied $0 \le \Omend \le \Omout$ for constant $\Omin=\Omout=100$,
similarly we considered $0 \le \Omin=\Omout \le 100$ for $\Omend=100$ to
get values of the Ekman/Hartmann blowing and suction when the difference
between cylinder and end-plates rotation is large.  The agreement
with previous nonlinear calculation for the infinite plate  is quite
good,~\citep{BentonChow72}.  The dependence of the induced mass flux
and the current on the strength of the magnetic field as well as on the
relative fluid/end-plate rotation has the same character.

If the flow is vertically bounded by two plates, as for the Taylor-Couette
system, three essentially different regions can be distinguished: the
Ekman-Hartmann layer, a magnetic diffusion region and a current-free
region~(\citep{LoperBenton70,BentonLoper69}).  In the magnetic diffusion
region~(MDR) the axial Hartmann current must be reduced to zero before
it reaches the current-free region and, by continuity, it is turned
into radial direction.  This radial perturbation current interacts with
axial magnetic field and results in accelerating (for negative $j_R$
and positive $B_z$) or decelerating (for positive $j_R$) electromagnetic
body force.

The MDR arises since the Ekman-Hartmann layer itself is incapable
to force the current to satisfy the exterior boundary conditions.
It constantly grows in time and it quickly dominates the whole space
between the plates.  Moreover, when considering the small $\Pm$ limit,
the MDR  instantly becomes spatially uniform and infinitely thick even
for one bounding plane and the current-free region does not exists at
all,~\citep{BentonLoper69}.

Consequently in our enclosed MHD Taylor-Couette system with  $\Pm
\rightarrow 0$ we have relatively thin Ekman-Hartmann layer close to
the plates, whereas the fluid in the major part of the container forms
the MDR in which the axial Hartmann current changes into radial one.
We underline here that this is true for perfectly conducting walls,
since such radial boundary conditions assure us that the current can
penetrate the cylinders.  The situation would be rather different with
insulating radial boundaries.

\subsection{\label{sec:conducting}Conducting end-plates}

For highly conducting plates the induced current drawn into/from
the plates is much stronger than the current induced in the layer for
insulating boundaries.  The Ekman-Hartmann layer itself is nearly 
unaffected by conductivity of the plates as are the velocities and the 
currents within this layer.  
However, due to constant magnetic field perturbation there exists 
an additional electric current of order
$2\Phi$ which is induced by the conducting boundaries, 
$ \Phi = \epslayerk \sqrt{\Omplain/\nu} $, and $\epslayerk$ characterizes the
relative conductance of the fluid and the thin plates,~\citep{Loper70}.
Moreover, in the MDR this current increases fluid 
velocity by a factor of $2 \alpha^2 \Phi$.

\begin{figure}[htb]    
 \mbox{\includegraphics[width=\stdwid]{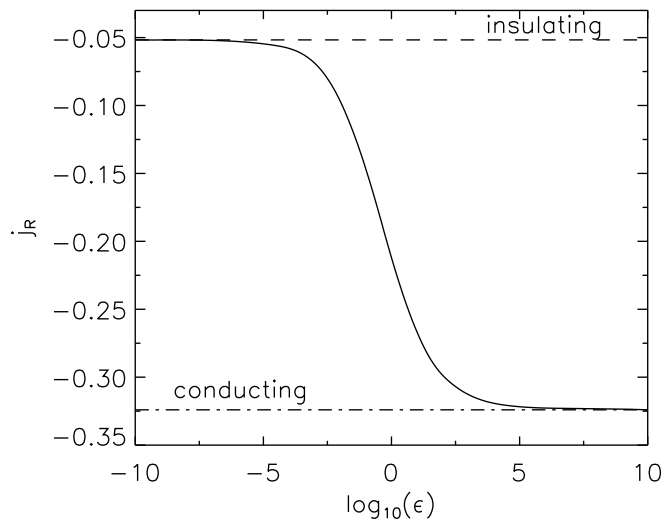}}
 \caption{ \label{fig:jrepslayer} 
    Radial current $j_R(R=\Rin/D+1/2, z=\Gamma/2)$ in the 
    middle of the gap for MHD Taylor-Couette flow 
    with $\Ha=10, \Omout=\Omin=200, \Omend=202, \Gamma=10$. 
    The upper line represents insulating plates,
    the bottom line perfectly conducting ones, and in between is the intermediate 
    case for different values of the relative conductance. } 
\end{figure}

Figure \ref{fig:jrepslayer} shows how the radial current $j_R$ in the
middle of the container changes with conductivity of the end-plates.
The difference between the perfect insulator and the perfect conductor
is almost one order of magnitude even for so slow rotation.  We find
that for an MRI experiment it is \emph{crucial to use insulating plates}
in order to minimize this undesirable current.

\section{The influence of the Hartmann current}

We notice that the force due to the radial current and the axial magnetic field, 
$\Ha^2 j_R \unitr \times \Bext/B_0$, enters the 
momentum equation \eqref{eq:momentumB} for $u_\phi$ component.  
Formally the force is equivalent to applying an azimuthal pressure gradient $\partial_\phi p \ne 0$.
A flow between rotating cylinders with non-zero $\partial_\phi p$ is usually 
referred as the Taylor-Dean flow,~\citep{Chandra60}. Its rotational profile $\Omegad$ is
a superposition of the circular Couette profile~\eqref{eq:couprofile}
and the steady flow
 \[
  \label{eq:tdbasic}
  \Omegad = \Omegab + e\left( c + d/R^2 + \ln R \right),
 \]
with 
 \[
   c&=&\frac{\Rin^2 \ln(\Rin) - \Rout^2 \ln(\Rout)}{\Rout^2 - \Rin^2}, \\
   d&=&\frac{ \Rin^2 \Rout^2 \ln (\Rout/\Rin)}{\Rout^2 - \Rin^2}, \\
   e&=&\frac{1}{\rho \nu} \pregrad.
 \]
The pressure gradient can be realized by an external pumping
mechanism or, like in the discussed case, by the Lorentz force
resulting from the induced current and the axial magnetic field. 

Let us introduce a parameter $\dpump$ describing Taylor-Dean flows, 
the ratio of average pumping velocity to the rotation velocity
\[
 \label{eq:dpumpdef}
 \dpump = \frac{6 V_m}{\Omin \Rin},
\]
where $V_m$ is the average pumping velocity 
\[
 \label{eq:vmdef}
 V_m &=& \frac{1}{D} \int_{\Rin}^{\Rout} \left[ e\left( c + d/R^2 + \ln R \right) \right] \mathrm{d}R \nonumber \\
     &=& -\partial_\phi p \frac{\Rout}{2 \rho \nu} \frac{ (1-\etah^2)^2 - 4\etah^2(\ln{\etah})^2 } { 4(1-\etah)(1-\etah^2) },
\]
\citep{Chen93}. The basic question arises whether the resulting 
pumping due to the radial current and the
axial field can bring the flow into an unstable regime.

\subsection{Hartmann current generated by the end-plates}

The structure of the Ekman-Hartmann layer changes with parameters such
as rotation rates or strength of the magnetic field.  Here, however,
we will concentrate on the flow in the bulk of the container so that
only currents and velocities which leave the layer are important.
We analyze hydrodynamically stable flow with $\muh=0.27$ at the aspect
ratio $\Gamma=10$ with rigidly rotating end-plates.

\subsubsection{End-plates rotating with $\Omout$ and $\Omin$}

First we consider cylinders covered with rigid, perfectly
conducting plates rotating with the angular velocity equal to that of
the outer cylinder $\Omend = \Omout$.  We choose conducting lids so that
the induced current is much stronger and its influence on the flow is
more evident.

 \begin{figure}[tbh]
    \mbox{ \def\wid{2cm} \includegraphics[width=\wid]{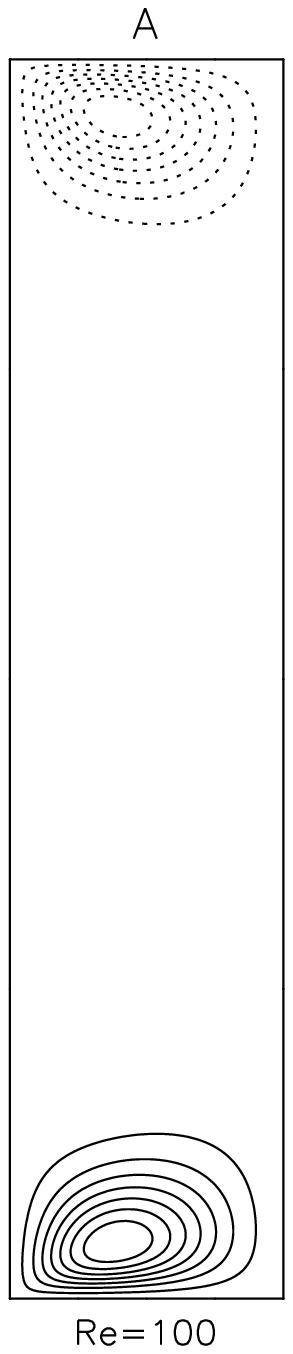}
    \includegraphics[width=\wid]{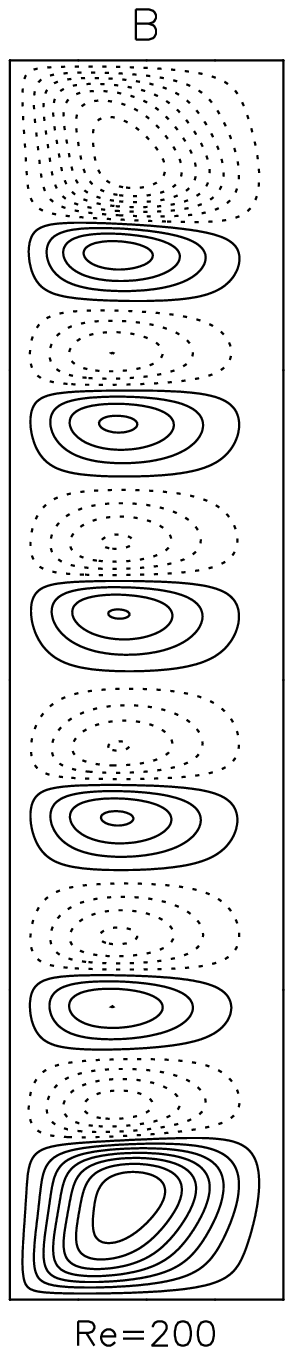}
    \includegraphics[width=\wid]{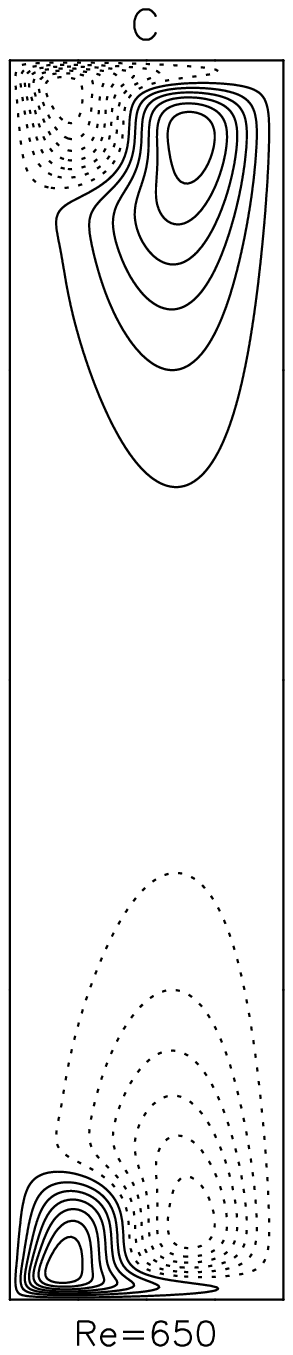}
    \includegraphics[width=\wid]{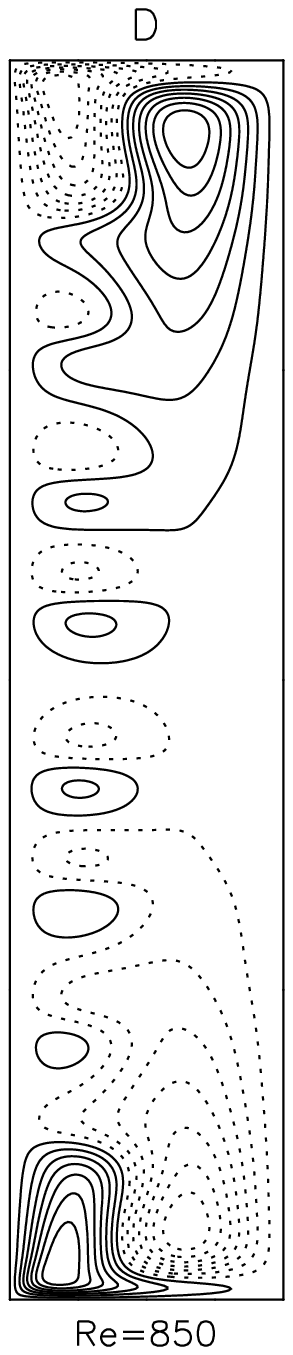} }
    \caption{ \label{fig:lettersAD}
     Contour lines of stream function for different flow parameters, the
     left edge of each panel denotes the inner cylinder, the right edge
     the outer one, solid lines correspond to clockwise fluid rotation.
     The end-plates are attached to the outer cylinder, $\Omend=\Omout$,
     $\muh=0.27, \Gamma=10$. ``A'', ``B'' are for conducing plates and
     $\Ha=3$, ``C'', ``D'' for insulating ones and $\Ha=10$. }
 \end{figure}

When the plates rotate with $\Omend = \Omout$, the Ekman circulation is
clockwise and the corresponding Hartman current has the positive sign,
i.e., close to the inner cylinder it leaves the Ekman-Hartmann layer
with $j_z > 0$, consequently the radial current also has positive sign.
Figure~\ref{fig:lettersAD}~A-B displays a flow with conducting plates and
a weak axial magnetic field applied, $\Ha=3$, for two different Reynolds
numbers. The rotation ratio is $\muh=0.27$, so that the Couette flow
is hydrodynamically \emph{stable}, however we notice that when $\Rey$
is large enough the flow changes significantly and the Taylor vortices
can be observed.

This phenomenon can be explained as follows: for a constant $\Ha$,
increasing of the rotation rate leads to the stronger Hartmann current
drawn into the flow, therefore the corresponding pumping $\dpump$ due
to $j_R \unitr \times \Bext$ increases and for certain $\Rey$ it reaches
a critical value $\dpumpc$, so that the instability develops.

If the perfectly conducting ends are replaced with insulating ones the
induced current is much weaker.  When the imposed magnetic field has
strength such that $\Ha=3$, the pumping is too small to make the flow
unstable, regardless of the Reynolds number.  However, when the magnetic
field is stronger, $\Ha=10$, for sufficiently high rotation rates the
vortices can also be seen, Fig.~\ref{fig:lettersAD}~C-D.

It is known that stronger axial magnetic field has a
stabilizing effect even on a hydrodynamically unstable
flow~\citep[e.g.][]{RudigerSchultzShalybkov03}.  Besides that, the
Hartmann current increases with the amplitude of the magnetic field
only until a certain point is reached.  When the magnetic interaction
parameter becomes $\alpha \approx 2.5$, increasing $\Ha$ does not further
increase the Hartmann current,~\citep{GilmanBenton68}.  For these reasons
it is clear that when the imposed magnetic field is strong enough the
instability described above will not occur.  Indeed, it has been checked
that for conducting plates, $\Rey=200$ and the magnetic field with
$\Ha=20$ there are no Taylor vortices, although the rotational profile
is significantly changed when compared to the non-magnetic situation.

\begin{figure}[tbh]
    \mbox{ \def\wid{2cm} \includegraphics[width=\wid]{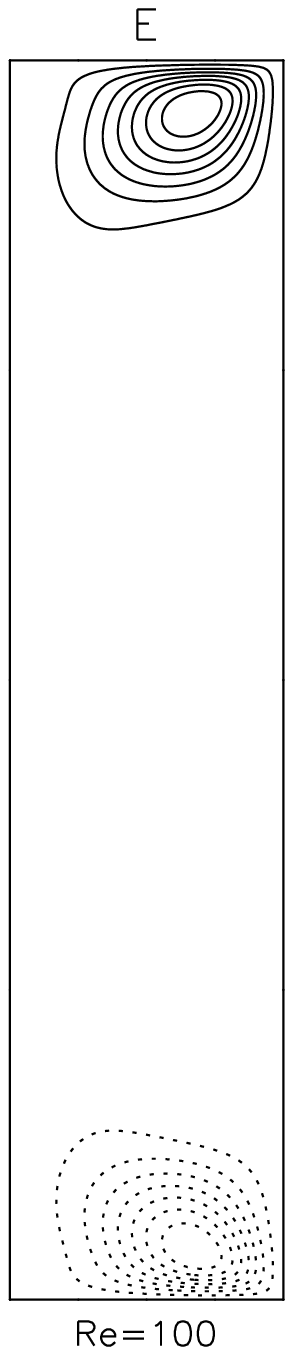}
    \includegraphics[width=\wid]{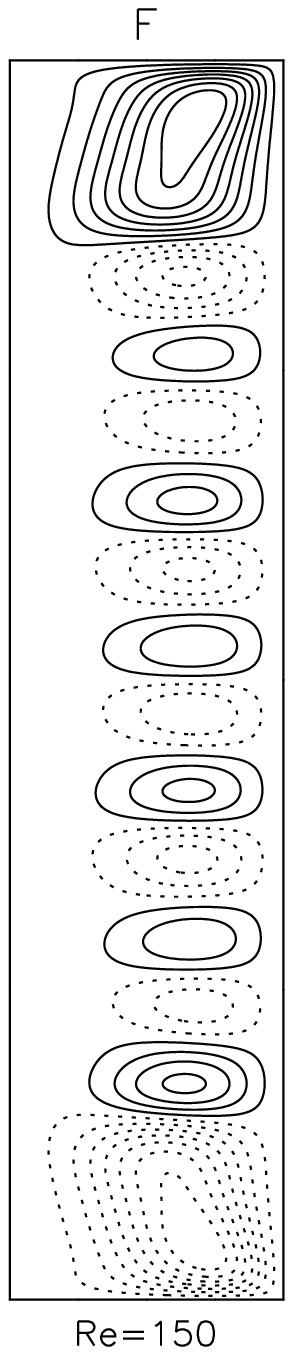}
    \includegraphics[width=\wid]{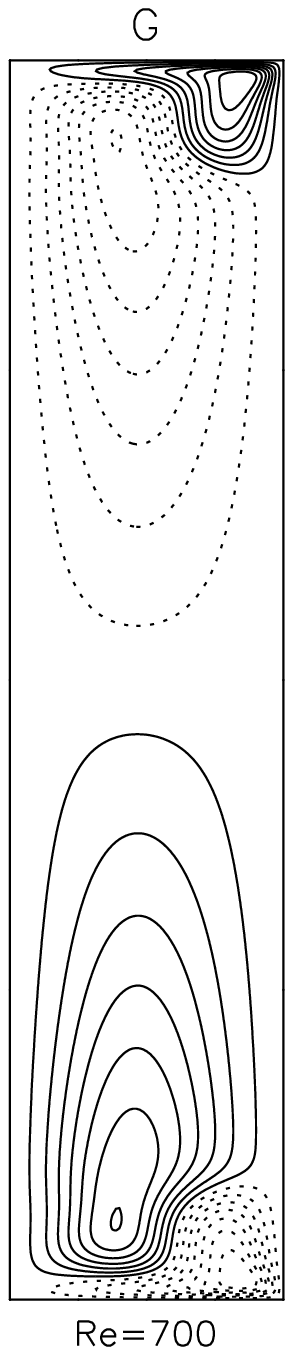}
    \includegraphics[width=\wid]{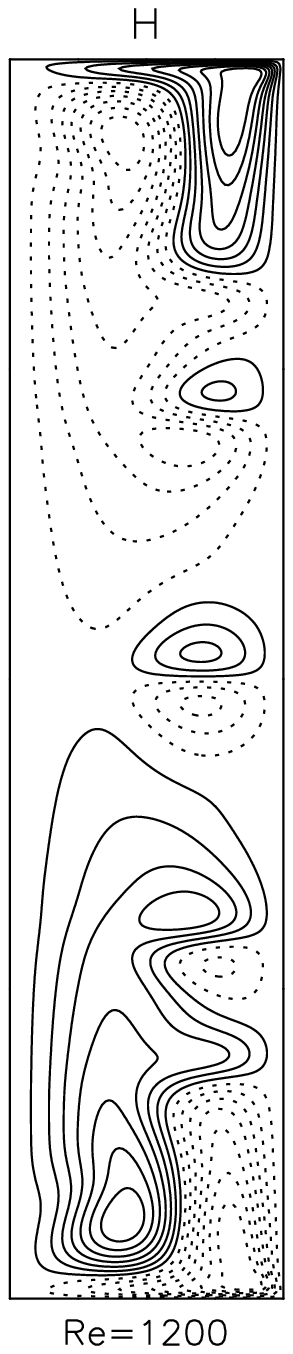} }
    \caption{ \label{fig:lettersEH}
     The end-plates are attached to the inner cylinder,
     $\Omend=\Omin$. ``E'', ``F'' are for conducing plates and $\Ha=3$,
     ``G'', ``H'' for insulating ones and $\Ha=10$.  Note that for the
     insulating endplates $\Rey$ is order of magnitude larger.}
 \end{figure}

If rigidly conducting end-plates are attached to the inner cylinder, so
that $\Omend=\Omin$, the Ekman circulation is counter-clockwise (Ekman
suction) and the corresponding Hartmann current has a negative sign, so
that the parameter $\dpump$ is positive.  From Fig.~\ref{fig:lettersEH}~E-F
we see that, analogously to the case ``C''-``D'',
if the rotation is sufficiently fast the resulting $\dpump$ reaches
critical value and the flow becomes dominated by the vortices.

Similarly when the insulating plates are used, the axial magnetic
field with $\Ha=3$ is too weak to generate sufficiently large $\dpump$.
When stronger field is applied, $\Ha=10$ it is possible to 
observe the instability (``G''-``H'').

\subsubsection{The rotational profile}

\begin{figure}[bth]    
 \mbox{ \includegraphics[width=\stdwid]{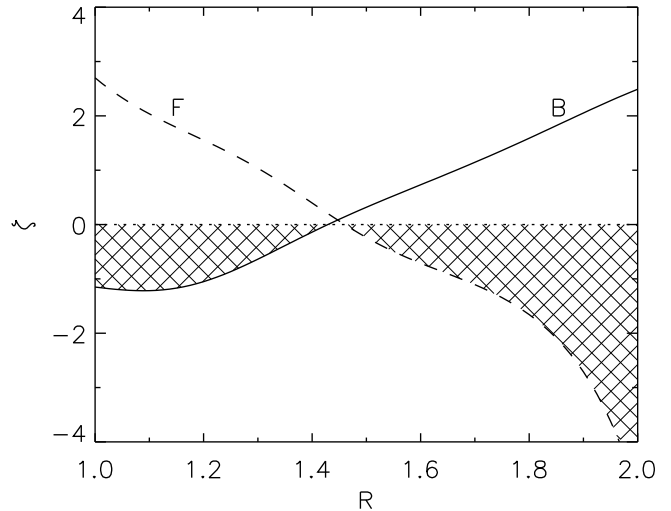} }
 \caption{ \label{fig:zetoman} 
    $\raydisc(R,z=\Gamma/2)$ calculated for MHD Taylor-Couette flow
    with conducting end-plates attached to the outer cylinder (``B'', $\Rey=200$) 
    and the inner one (``F'', $\Rey=150$).  The Rayleigh
    criterion yields $\raydisc > 0$ for stability.
    } 
\end{figure}

As mentioned above, if the induced radial current has the same sign as the 
axial magnetic field, the azimuthal velocity of the fluid is decelerated, if the signs
are opposite the flow is accelerated. 
The discussed instability is a centrifugal one and is simply due
to change in rotational profile of the fluid.  
Let us use a Rayleigh
discriminant for stability, $\raydisc = \partial_R (R^2 \Omega) / (R \Omega)$, 
the flow is stable if $\raydisc > 0$.  Fig.~\ref{fig:zetoman} shows
the radial dependence of $\raydisc$ in the middle of the gap ($z=\Gamma/2$) for
the two cases labeled as ``B'' and ``F''. 

We notice that the vortices concentrate in region where $\raydisc$
is negative, i.e., where the Rayleigh criterion is not fulfilled.
This instability has essentially local character, so it is not possible
to define any specific critical Reynolds number whose crossing would lead
to some exponential grow in the whole container.  For the conducting
plates and $\Omend=\Omout$, there exists $\Rey$ between 100 (``A'')
and 200 (``B'') for which only a part of the container would be filled
with the vortices.

\subsection{\label{sec:linstab}Linear stability of current-induced MHD Taylor-Dean flow}

In order to predict the onset of the instability discussed above we
analyze the global stability of MHD Taylor-Dean flow for our parameters.
For the nonlinear simulations we can estimate the pumping due to the
azimuthal pressure gradient just by setting $(\nabla p)_\phi$ to $\Ha^2 j_R$, 
see Eq.~\eqref{eq:momentumB}.  Generally $(\nabla p)_\phi$ and $j_R$
change with radius like $R^{-1}$.  However, due to the presence of the plates, 
for $j_R$ this is true only far from the vertical boundaries and here 
the value of $j_R$ is taken at $R=\Rin/D, z=\Gamma/2$ (note that for our 
perfectly conducting boundaries the current penetrates the cylinders and 
for a steady state it is largest at $R=\Rin/D$).  
In this way we obtain the
parameter $\dpump$ associated with the enclosed MHD Taylor-Couette for the
given boundary conditions ``A''-``H'', and then it can be compared with
a critical value $\dpumpc$ obtained from the linear stability analysis.

Consider now the axisymmetric MHD Taylor-Dean flow for infinitely long
cylinders governed by the Eqs.~\eqref{eq:mhd}.  It admits the basic
solution $u_\phi = R \Omegad$ with $u_R = u_z = b_R = b_\phi = 0$ and
the imposed axial magnetic field $B_0$. The perturbed state is 
$u_R', R \Omegad + u_\phi', u_z', b_R', b_\phi', B_0 + b_z.$

After developing disturbances into normal modes we seek
solutions of the linearized MHD equations in the form similar like
in~\citep{RudigerSchultzShalybkov03, RudHall04}.  An appropriate set of
ten boundary conditions is needed in order to solve the system, these are
the no-slip boundary conditions for the velocity $u'_R=u'_\phi=u'_z=0$
and perfectly conducting for the magnetic field 
$\partial_R b'_\phi + b'_\phi/R = b'_R = 0$ at the both cylinders.  
We will only consider stationary marginally stable modes.

The homogeneous set of equations together with the boundary conditions
for the walls determine an eigenvalue problem of the form 
$L(\muh, \etah, k, m, \Pm, \Rey, \Ha, \dpump) = 0$.  The variables are approximated
with finite difference method on a grid typically with 200 points. The
numerical code used to solve the problem is identical to that used
in~\citep{RudigerSchultzShalybkov03}

For the current axisymmetric (see, however, \footnote{It should be mentioned that
for a Taylor-Dean flow,  it is fairly easy to excite oscillatory,
non-axisymmetric modes~\citep[see e.g.][]{Chen93}.}) study we set parameters $m=0,
\etah=0.5, \muh=0.27, \Pm=10^{-6}$, then for given $\Ha$ and $\Rey$ we
look for minimal value of $|\dpump|$ leading to the instability
(the value for which the determinant $L$ is zero).  Since $\dpump$
is directly proportional to the azimuthal pressure gradient, and therefore
to the radial current, the resulting critical $\dpumpc$ determines
the minimum value of the radial current for which the Taylor-Dean flow
becomes unstable.

\begin{figure*}[hbt]
    \mbox{  \includegraphics[width=\stdwid]{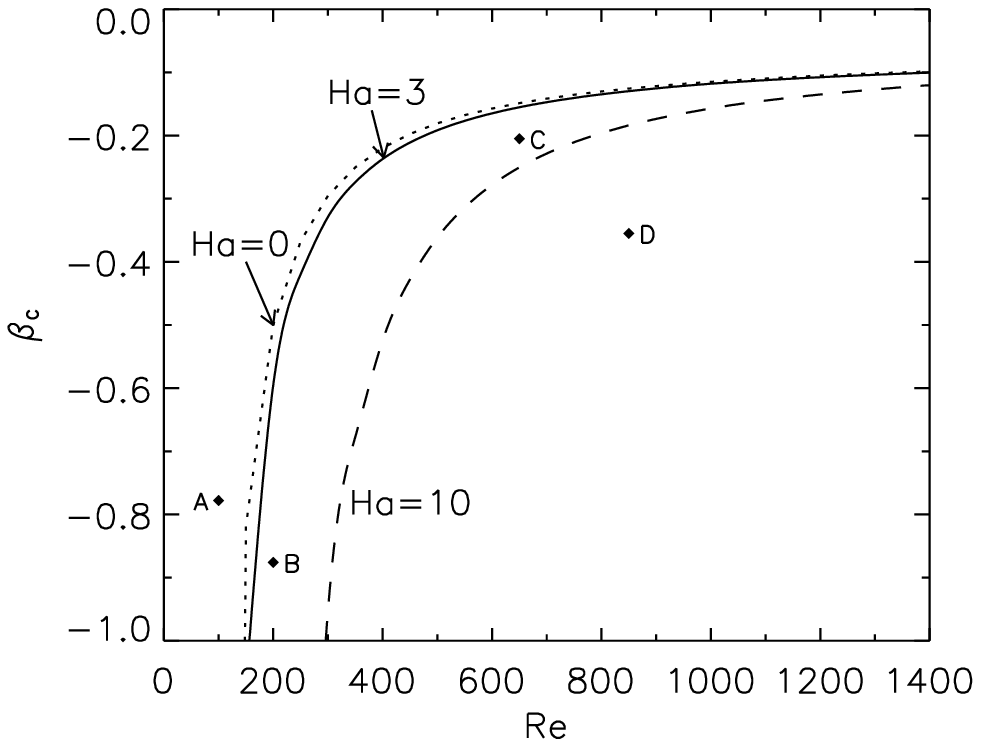}
    \includegraphics[width=\stdwid]{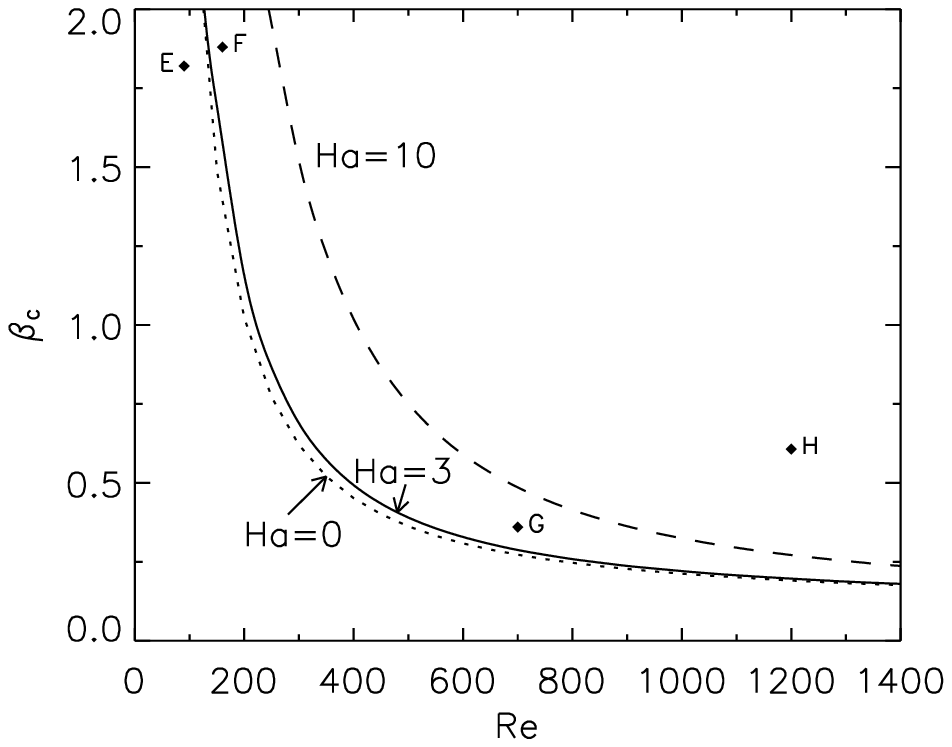} }
    \caption{ \label{fig:jrstabil}
    Critical values of the pumping/rotation ratio $\dpumpc$ for
    MHD Taylor-Dean flow with $\muh=0.27$, for different Reynolds
    number $\Rey$ and strength of the magnetic field $\Ha$.
    The case for negative (positvie) $\dpump$ corresponds to the
    azimuthal pressure gradient due to positive (negative) radial
    currents interacting with the axial field.  On the left panel
    the unstabel region lies below the line, on the right one above
    the lines.  The letters ``A''-``H'' represent states displayed in
    Figs.~\ref{fig:lettersAD},~\ref{fig:lettersEH}.
    }
\end{figure*}

Figure~\ref{fig:jrstabil} shows marginal stability lines for the MHD
Taylor-Dean flow for different values of the imposed axial magnetic field,
for both positive and negative values of $\dpump$.  We notice that much
larger values of $|\dpump|$ are needed for stronger axial magnetic fields
since the field plays a stabilizing role.

The labels ``A''-``H'' refer to MHD Taylor-Coutte flows presented in
the previous section.  E.g., ``A'' refers to the flow with $\Rey=100, \Ha=3, \muh=0.27$ 
with perfectly conducting end-plates attached
to the outer cylinder.  The induced current $j_R$ is such that the
corresponding $\dpump$ due to the Lorentz force denotes a stable flow.
If the Reynolds number is increased, the critical value $\dpumpc$
(for $\Ha=3$) is reached and the instability develops -- label ``B''.

\section{Summary}

\citet{GilmanBenton68} have shown with a linear theory that in vicinity
of a rotating plane which serves as a border for rotating conducing
fluid there develops the Ekman-Hartmann layer if $\Omplain \ne \Omfluid$
and an axial magnetic field is applied.  The most important feature of
Ekman-Hartmann layers is their ability to induce both mass fluxes and
electric currents in the region outside the boundary layer.  If 
$\Omplain < \Omfluid$ these fluxes are directed outwards the layer (``blowing'');
when $\Omplain > \Omfluid$ towards the layer (``suction''). For the
conducting plates the fluxes are much stronger since additional currents
are drawn from/into the plates.

Outside the Ekman-Hartmann layer exists the magnetic diffusion region,
in which the electric current has only radial components. The current,
together with the axial magnetic field, produces an electromagnetic body
force acting on the fluid.

We have shown in this paper that similar effects arise for the MHD
Taylor-Couette flow when the rotating cylinders are bounded by two rigidly
rotating end-plates.  Near the plates the Ekman-Hartmann layer forms and,
consequently, there exists the Hartmann current which penetrates bulk of
the fluid. In the presence of an axial magnetic field such problem can be
compared with the Taylor-Dean flow -- a flow between, possibly rotating,
cylinders which is additionally driven by an azimuthal pressure gradient.

We find that under certain conditions the resulting flow becomes
unstable, Taylor vortices can be observed and the rotational profile is
significantly different from the standard Couette solution $\Omegab$.
The instability has essentially a centrifugal character as the Rayleigh
criterion is locally vialoted.  This is an undesirable effect from the
point of view of an MRI experiment.  In such experiment it is necessary to
obtain a state resembling $\Omegab$ in the major part of the container,
for parameters characterizing stable MHD flows.  It is necessary to take
into account the magnetic effects induced by the plates so that the MRI
can be clearly identified rather then any other instability.

The fluxes induced in the Ekman-Hartmann layer are a direct consequence
of a shear close to the boundaries.  Exemplary methods of reducing the
shear have been proposed in~\citep{Szklarski07}.  For rotation rates
characterized by $\Rey$ of order \ord{10^3} all the effects can by
significantly reduced by allowing the end-plates to rotate independently
of the cylinders~\citep{Abshagenetal04}.  Since for $\Omend = \Omin$
there is the Ekman suction, and for $\Omend=\Omout$ the Ekman blowing,
there exists $\Omout < \Omend < \Omin$ for which the generated mass and
charge fluxes are minimal.  Alternatively, one can divide the plates
into independently rotating rings~\citep{JiBurin06}.



\end{document}